\documentclass[aps,pre,preprint,amsmath,amssymb,groupedaddress,showkeys]{revtex4}
\usepackage{amsmath,amssymb}
\usepackage{epsfig}
\usepackage{graphics}
\usepackage{graphicx}
\usepackage{float}
\usepackage{dcolumn}
\usepackage{xcolor}
\begin{document}

\title{An experimental analogue for memristor based Chua's circuit}

\author{R.~Jothimurugan}
\email[]{jothi@cnld.bdu.ac.in}
\affiliation{Centre for Nonlinear Dynamics, School of Physics, Bharathidasan University, Tiruchirappalli-620 024, Tamilnadu, India}
\author{Ronilson Rocha}
\email[]{rocha@em.ufop.br}
\affiliation{Federal University of Ouro Preto-UFOP/EM/DECAT, Campus Morro do Cruzeiro, 35400-000, Ouro Preto, MG, Brazil.}
\author{K.~Thamilmaran}
\email[]{maran.cnld@gmail.com}
\affiliation{Centre for Nonlinear Dynamics, School of Physics, Bharathidasan University, Tiruchirappalli-620 024, Tamilnadu, India}
\date{\today}

\begin{abstract}
This paper presents the dynamics of a practical equivalent of a smooth memristor oscillator derived from Chua's circuit. This approach replaces the conventional Chua's diode by a flux controlled memristor with negative conductance. The central idea is to design a practical memristor based circuit using electronic analogy in order to bypass problems related to the realization of memristor equivalents. The amplitude and the frequency of the oscillations are previously defined in the circuit design. The result is a robust and flexible circuit without inductors, which is able to reproduce a rich variety of dynamical behaviors. The proposed analogue circuit is successfully designed and implemented, producing experimental time series, phase portraits, and power spectra, which are corroborated by numerical simulations. The ``0-1 test" is also performed in order to verify the regular and chaotic dynamics on the proposed analogue circuit.	
\end{abstract}


\maketitle

\section{Introduction}
Leon O. Chua postulated the memristor in the year of 1971 based on the symmetry arguments. It is believed to be the fourth fundamental passive circuit element followed by the resistor, capacitor and inductor \cite{ref1}.  Memristor gets wide attention by the research community after its physical realization by Stanley William's group from HP labs in 2008 \cite{ref2}. The memristor is a passive two-terminal electronic device described by a nonlinear constitutive relation which is either one of the following two forms of the voltage-current relationship, (i) $v=M(q)i$ and (ii) $i=W(\phi)v$. Here, the nonlinear functions $M(q)$ and $W(\phi)$ are regarded as {\it{memristance}} and {\it{memductance}} respectively, whose functional forms are defined by $M(q)=d\phi(q)/dq$ and $W(\phi)=dq(\phi)/d\phi$ represent the slope of the scalar functions $\phi=\phi(q)$ and $q=q(\phi)$, respectively which are called memristor constitutive  relation \cite{ref3}.

The memristor has potential applications in physics, mathematics, engineering, biology, and other various fields. Recent researches that involve synthesis of memristive devices \cite{ref2,ref4}, mathematical modeling and analysis \cite{ref5,ref6,ref7,ref8}, application of memristor theory to the nano-battery \cite{ref9}, memory effects on light emitting diodes \cite{ref10}, memristive behavior of electrical properties of the human skin \cite{ref11}, and synaptic connections among brain cell of the human \cite{ref12} explicitly demonstrate the potentiality of the memristor. In a nutshell, it can be understood from the 18K+ citations registered for memristor in google scholar as on 01.07.2014 \cite{ref13}.   

The analysis of nonlinear systems using experimental circuits is always an active area of research that provides a better understanding on the theoretical concepts. The FitzHugh-Nagumo circuit for neuron model \cite{ref14,ref15}, the Lorenz oscillator based on a climate model \cite{ref16} and the popular Chua's circuit \cite{ref17} are few examples of this kind. In this context, several studies are made on the development of memristor based circuits, since memristors are commercially not available yet. For example, the analog integrator based smooth memristor was implemented to study the behaviour of Chua's circuit with memristor as nonlinear element \cite{ref18}. The time varying resistor based memristor circuit is implemented on electronic circuit simulator to study the occurrence of chaotic beats in a driven Chua's circuit \cite{ref19} and the nonsmooth bifurcations, transient hyperchaos and hyperchaotic beats in a Murali-Lakshmanan-Chua circuit \cite{ref20}.  Apart from that, analogue circuits to explore the voltage-current characteristic of the memristor are also reported in the literature \cite{ref21,ref22}. In all these studies, the memristor is separately conceived as a sub-circuit to follow the specific choice of memristive function. All these sub-circuits are complex and prone to the dynamical deviations from the mathematical modeling due to the tolerance effects of the circuit components and the inherent noise effects. Further, all these reports having physical inductors which are unwise to implement the circuits on PCB/VLSI designs.

In this paper, the smooth memristor based Chua's circuit equation is implemented in an experimental hardware in order  to capture its dynamical behaviors on real time. The circuit is designed using the concept of electronic analogy \cite{ref23,ref24}. The advantage of this realization is that the amplitude of the each state variable can be scaled up/down according to the application requirements, reducing the noise influence on the circuit dynamics. This method of circuit implementation is straight forward and easy to implement with printed circuit boards (PCBs) since it do not have physical or synthetic inductors. The functional form of the memristor associated with this circuit is assumed to be smooth, continuous and monotonically increasing nature. The proposed analogue circuit realization is versatile, stable, and it satisfactorily reproduces almost all simulated dynamics by numerical integration. The experimental time series, phase portraits and power spectra are traced in real time on digital storage oscilloscope, which are confirmed using numerical simulation. We also have performed the newly developed ``0-1" test to the experimental times series to distinguish the regular and chaotic motion of the circuit dynamics. The parameter $K$ results `0' for regular and `1' for chaotic dynamics.

The rest of the paper is organized as follows: the original memristor based Chua's circuit, its state equations and the associated dynamics are discussed in Section 2. In Section 3, first the procedure of amplitude scaling is explained and then, the method of constructing Chua's circuit using electronic analogy is described. The results of PSpice simulation, experimental observation and the numerical integration studies are discussed in Section 4. In Section 5, the ``0-1 test" is performed using experimental time series data to verify the periodic and chaotic dynamics of the circuit.  Conclusions are given in section 6.
\section{The Chua's circuit with Memristor}
The Figure 1 shows the memristor based Chua's circuit. Here, $N_R$ is active nonlinear element which comprises the parallel combination of negative conductance ($-G$) and two terminal passive flux controlled memristor. The memristor is characterized by smooth continuous cubic monotonic-increasing nonlinearity given by \cite{ref3,ref18}:

\begin{equation}
q(\phi)=a\phi+b\phi^3,
\end{equation}

{\noindent where $a$ and $b$ are constants which are assumed to be greater than zero. From this expression, the memductance $W(\phi)$ can be deduced as}

\begin{equation}
W(\phi)=\frac{dq(\phi)}{d\phi}=a+3b\phi^2.
\end{equation}

The circuit equation for memristor based Chua's circuit can be written using Kirchoff's laws as follows \cite{ref3},

\begin{subequations}
\begin{eqnarray}
\frac{dv_1}{dt}&=&\frac{1}{RC_1}(v_2-v_1+GRv_1-RW(\phi)v_1),\\
\frac{dv_2}{dt}&=&\frac{1}{RC_2}(v_1-v_2-Ri_L),\\
\frac{di_L}{dt}&=&-\frac{1}{L}v_2,\\
\frac{d\phi}{dt}&=&v_1,
\end{eqnarray}

{\noindent where, $v_1$, $v_2$, $i$ and $\phi$ are voltage across the capacitors $C_1$, $C_2$, current flowing through the inductor $L$ and the flux through the memristor respectively. Substituting the functional form $W(\phi)$ in to Eq. 3(a) yields,}

\begin{equation}
\frac{dv_1}{dt}=\frac{1}{RC_1}(v_2-v_1+GRv_1-R(a+3b\phi^2)v_1).
\end{equation}
\end{subequations}
By applying transformation to the variables as $x_1=v_1$, $x_2=v_2$, $x_3=i_L$, $x_4=\phi$, $\alpha=1/C_1$, $\beta=1/L$, $\zeta=G$, $C_2=1$ and $R=1$, then the circuit equations becomes,
\begin{eqnarray}
\begin{aligned}
\frac{dx_1}{dt}&=\alpha(x_2-x_1+\zeta x_1-(a+bx_4^2)x_1),\\
\frac{dx_2}{dt}&=x_1-x_2-x_3,\\
\frac{dx_3}{dt}&=-\beta x_2,\\
\frac{dx_4}{dt}&=x_1.
\end{aligned}
\end{eqnarray}

{\noindent The dynamics of Eq. (4) is obtained through numerical simulation. Throughout our analysis, Runge-kutta fourth order algorithm is  used to solve the equations numerically. For the following specific choice of system parameters in the Eq. (4), $\alpha=9.8$, $\beta=100/7$, $\zeta=9/7$, $a=1/7$, $b=2/7$ and initial conditions \{$x_1, x_2, x_3, x_4$\}=\{$0,0,0.1,0$\}, the chaotic motion is obtained. Figure 2 shows the phase portraits of the system in different projections of $x_1, x_2, x_3$ and $x_4$ planes.} 
 
\section{Electronic analogue of memristor based Chua's circuit}

\subsection{Scaling up the circuit variables}
The maximum amplitude of oscillation of variables $x_1$, $x_2$, $x_3$ and $x_4$ of Eq. (4) are obtained from the simulated time series of the chaotic motion after vanishing the initial transients. They are $|x_1|=1.3965$ V, $|x_2|=0.4772$ V, $|x_3|=2.3797$ A and $|x_4|=0.9441$ Wb. The direct implementation of memristor based Chua's circuit for the chosen parameter set in Eq. (4) is not possible due to the current $i_L$ or $x_3$ is 2.3 A, which is well above the current rating of the usual electronic circuit design. Also, the variable $x_4$ is in the units of flux which is not a physically measurable quantity. So the analagoue circuit simulation is the suitable method for this circuit implementation. In the analogue circuitry, all variables are converted into voltages. Since the maximum amplitudes of the state variables $x_2$ and $x_4$ are less than $1$, the signals of these variables are more prone to affected by the inherent noise while implementing the circuit. To avoid these limitations, we uniformly upscale the maximum value of all the variables to $\pm 5$ V. This range is adequate to avoid noise and saturation effects over the op-amps considering a power supply of $\pm 9$ V. Further, it is convenient to visualize the time waveform in the oscilloscope traces as well as to capture the data through data acquisition cards. Applying the amplitude scaling to the Eq. 4, the new variables are defined as $\overline{x}_1=3.3x_1$, $\overline{x}_2=10x_2$, $\overline{x}_3=2x_3$ and $\overline{x}_4=5x_4$. Rewriting the Eq. (4) to the new variables and dropping the bar over the variables for convenience, yields the following upscaled system,

\begin{eqnarray}
\begin{aligned}
\frac{dx_1}{dt}&=\alpha(0.33x_2-x_1+\zeta x_1-ax_1+0.12bx_4^2x_1),\\
\frac{dx_2}{dt}&=3.03x_1-x_2-5x_3,\\
\frac{dx_3}{dt}&=-0.2\beta x_2,\\
\frac{dx_4}{dt}&=1.5152x_1.
\end{aligned}
\end{eqnarray}

Considering a probability of any intermediate signal to surpass the saturation voltage limits, the entire system equation is then divided by the largest parameter. In the present case, since the greatest parameter is $\beta=100/7=14.28$, we divide the entire system by the factor 15 on both sides of the Eq. (5). Although this procedure affects the operational frequency of the circuit, it has not influence on the system dynamics. Hence, the rescalled final set of equations are given by,  

\begin{eqnarray}
\begin{aligned}
\frac{dx_1}{dt}&=0.2156x_2+0.09342x_1-2.24109\frac{x_4^2}{100}x_1,\\
\frac{dx_2}{dt}&=0.202x_1-0.067x_2-0.33x_3,\\
\frac{dx_3}{dt}&=-0.0133\beta x_2\\
\frac{dx_4}{dt}&=0.10101x_1.
\end{aligned}
\end{eqnarray}

In this rescaling, the parameter values used for numerical simulation of Eq. (4) is applied for all the parameters except $\beta$, which is used to be the control parameter for the rest of the study.

\subsection{Implementation of analogue Memristor based Chua's circuit}
In the analogue computation, a first order differential equation can be solved primarily using an weighted integrator. The dynamics of the memristor based Chua's circuit is described by a set of four first order coupled nonlinear differential equations as given in Eq. (6). Hence, the analogue implementation of memristor based Chua's circuit will have four weighted integrators. Unit gain inverting amplifiers are used to change the sign of the variables. To incorporate the nonlinear functional part present in the Eq. 6, the analogue multipliers are used. Figure 3 shows these three fundamental cells namely (a) the weighted inverting integrator, (b) scale changer (unit gain inverting amplifier) and (c) the multiplier chip, used to realize the analogue memristor based Chua's circuit. The transfer function of the weighted integrator is given by
\begin{equation}
v_o=-\frac{1}{RC}\int \! \frac{v_{in}}{R^*} \, \mathrm{d}t,
\end{equation}
where $v_{in}$ and $v_0$ are the input and output voltages of the integrator respectively. The $R^*$ refers to the normalized value of the input resistance in relation to the base resistance $R$. The frequency of the analogue circuit is determined by the value of the base resistance $R$ and the capacitance $C$. The sign of the any state variable can be electronically changed using unit gain inverting amplifier. The transfer function of the inverting amplifier is,
\begin{equation}
v_o=-v_{in}, ~~~~~~~~~~~{\mathrm{when}}~~~ R_{in}=R_f, 
\end{equation}
where $R_{in}$ and $R_f$ are the input and feedback resistors of the amplifier unit. The analog multiplier is useful component to multiply two analog signals. We use this component to get the nonlinear part present in the Eq. 6. The product output $W_0$ of a typical multiplier can be written as,

\begin{equation}
W_0=\frac{(X_1-X_2)*(Y_1-Y_2)}{10}+Z,
\end{equation}
Here, $X_1$ and $X_2$ are $X$-{\it multiplicand} of non-inverting and inverting inputs, $Y_1$ and $Y_2$ are $Y$-{\it multiplicand} of non-inverting and inverting inputs and $Z$ is the summing input. The dividing factor 10 is used to avoid the overflow of product output. 

The analogue circuit to emulate the Eq. (6) is implemented using the above described three electronic entities as shown in Fig. 4. In the circuit diagram $U1$, $U2$, $U3$ and $U4$ are the weighted integrators, $U5$ and $U6$ are inverting amplifiers and $U7$ and $U8$ represents the multiplier chips. A feedback resistor with high value ($R$ = $2$ M${\mathrm{\Omega}}$) is connected to each integrator in order to fix the gain at low frequencies and to reduce the effect of the offset voltages in the op-amps. This feedback resistor should be at least 10 times greater than the input resistance of its respective integrator. For similar reasons, resistors are placed at the $Z$ terminals of the multipliers also. 

The coefficients values values of the Eq. (6) are inversely proportional to $R^{*}$ values in the circuit. To explain the calculation of the $R^{*}$ value from the coefficients of Eq. (6), we consider last line in the Eq. (6). The variable $x_1$ in the right hand side has the coefficient value 0.10101. The corresponding $R^{*}$ value to the input resistance of the $W$ cell is $1/0.10101=9.9$. The value of corresponding resistor in the circuit is then directly calculated as ($R^{*}$x$R$) 9.9 x 10 k${\mathrm{\Omega}}$ = 99 k${\mathrm{\Omega}}$. In the similar way, values of the resistance corresponding to all coefficients in Eqns. (6) are obtained.

\section{Results}
\subsection{Experimental results}
On the implementation of analogous circuit, it is noticed that the calculated value of the resistances are not fit with the off-the-shelf component values. Thus, the direct implementation of this circuit shown in Fig. 4 is firstly realized on the PSpice, a popular electronic circuit simulation software. The component values given in Fig. 4 are used in PSpice simulation those are exactly the analogous to the numerical values. The double band chaotic attractor on different phase projections obtained from the PSpice simulation are given in Fig. 5 for $R15$ = 53.5 k${\mathrm{\Omega}}$.  Figure 6a shows the Fast Fourier Transform spctrum of the variable `$v_2$'. The measured central freqency in the FFT spectrum is 523 Hz. When the resistances are adjusted for off-the-shelf values considering the hardware realization, the circuit dynamics remains unaltered. The PCB hardware circuit is shown in Fig. 7. The values of the components are chosen as $C1=C2=C3=C4$ = 4.7~nF, $R1=R2=R3=R4$ = 2~M${\mathrm{\Omega}}$, $R5=R6=R7=R8$ = 10~k${\mathrm{\Omega}}$, $R9$ = 106.8~k${\mathrm{\Omega}}$, ($R9A$ = 100~k${\mathrm{\Omega}}$ + $R9B$ = 6.8~k${\mathrm{\Omega}}$), $R10$ = 100~k${\mathrm{\Omega}}$, $R11$ = 4.46~k${\mathrm{\Omega}}$, $R12$ = 47~k${\mathrm{\Omega}}$, $R13$ = 49.2~k${\mathrm{\Omega}}$, ($R13A$ = 47~k${\mathrm{\Omega}}$ + $R13B$ = 2.2~k${\mathrm{\Omega}}$), $R14$ = 150~k${\mathrm{\Omega}}$, $R16$ = 30~k${\mathrm{\Omega}}$, ($R16A$ = 10~k${\mathrm{\Omega}}$ + $R16B$=10~k${\mathrm{\Omega}}$ + $R16C$ = 10~k${\mathrm{\Omega}}$) and $R17=R18=~$6.8~k${\mathrm{\Omega}}$. These components may have the tolerance of 1\% from the standard value. To get the accurate reproduction of numerical results on experiment, the resistors $R9$, $R13$, and $R16$ are split into parts whereby the sum of the resistors will yield the closest possible value towards the numerical parameters. The variable resistor ``$R15$" will serve as the control parameter which is corresponding to $\beta$ used as control parameter in numerical analysis. The polyester type capacitors, TL081 op-amps and AD633JN multiplier chips biased with $\pm~$ 9 V dual power supply are used for this implementation. In Figs. (4) and (7), the markers $X$, $Y$, $Z$ and $W$ are the output terminals of the four variables $x_1$, $x_2$, $x_3$ and $x_4$ which we name as $v_1$, $v_2$, $v_3$ and $v_4$. The unit of these variables are obviously in `volts'. By varying the value of $R15$, different dynamical states of the circuit are captured on `Agilent 6014A DSO' oscilloscope. For instance, when $R15 = $15.6 k${\mathrm{\Omega}}$, the circuit generates the double band chaotic attractor on different projections in the phase plane as shown in Fig. 8. The corresponding time waveforms are plotted in Fig. 9. It clearly shows that all the variables are oscillating in the range of $\pm 5~$ V as defined in the circuit design. The central frequency of the variable $v_2$ is 454 Hz, which is measured from the power spectrum plotted in Fig. 6b. The dynamics of the circuit for the different values of the control parameter $R15$ in the range of 60 k${\mathrm{\Omega}}$ to 50 k${\mathrm{\Omega}}$ are given in Fig. 10, where is observed a clear evidence that this circuit presents the familiar period-doubling bifurcation route to chaos. 
 
\subsection{Numerical results}
In order to understand the complete dynamics of the circuit, Eq. (6) are numerically simulated.  The local maxima of oscillations in the variable `$x_4$' is considered for plotting one parameter bifurcation plot in the ($\beta-x_4$) plane, which is shown in Fig. 11 for the range of $12<\beta<20$. Further, the chaotic nature of the systems is quantified by calculating the Lyapunov exponent spectrum. The first three largest Lyapunov exponents ($\lambda_i$ for $i=1,2,3$) in the ($\beta-\lambda_i$) plane are plotted in Fig. 12. The presence of one positive Lyapunov exponent confirms the chaotic nature of the system.  The phase portrait of chaotic oscillations in different projections for $\beta=100/7$ is also shown in Fig. 13 , which is in agreement with the experimental phase portraits presented in in Fig. 8. The phase portraits during the period doubling transitions to chaos for different values of $\beta$ are shown in Fig. 14. Most of the plots shown in Fig. 14 is also captured on experiments as shown in Fig. 10. The numerical power spectrum with center frequency of 532 Hz is also plotted in Fig. 6c, which qualitatively matches well with the experimental spectra. The differences between simulated and experimental data can be mainly attributed to the tolerance of the real components and the presence of parasitic elements in the experimental set up.

\section{`0-1' Test}
Recently, a new test method has been proposed to characterize the regular and chaotic nature of the given data of deterministic system regardless of its dimensions \cite{ref25,ref26}. In this method the input is simply the time series of the system and the output will be either `0' for regular or `1' for chaotic motion of the system, hence it is named as ``0-1 test". It is applied to analyze electronic circuits \cite{ref27}, plasma \cite{ref28}, stock market \cite{ref29} and road transport \cite{ref30}. Consider a set of discrete one dimensional data $\Phi(n)$ sampled at the interval $n=1,2,3,...\Phi(n)$. Define the translation components $p(n)$ and $s(n)$ as  
\begin{equation}
p(n)=\sum^{n}_{j=1}{\Phi(j)cos(\theta(j)),}~~~~~~~n=1,2,3,..... 
\end{equation}
\begin{equation}
s(n)=\sum^{n}_{j=1}{\Phi(j)sin(\theta(j)),}~~~~~~~n=1,2,3,.....
\end{equation}
where
\begin{equation}
\theta(j)=jc+\sum^{j}_{i=1}{\Phi(i),}~~~~~~~~j=1,2,3,......,n.
\end{equation}

Here, the value of $c$ is constant and to be chosen arbitrarily. The dynamics of translation components (p, s) is regular in the ($p-s$) plane for regular and Brownian like motion for chaotic nature of the given data set. The plot of ($p-s$) plane itself gives an visual confirmation to the type of the motion for a chosen data set possess. Further, mean square displacement $M(n)$ is defined on the basis of $p(n)$ ($s(n)$ also would yield the same results) as
\begin{equation}
M(n)=\lim \limits_{N \to \infty}\frac{1}{N}\sum^{N}_{j=1}~[p(j+n)-p(j)]^2. ~~~~~n=1,2,3,....
\end{equation}

For the regular motion of $p(n)$, the $M(n)$ is bounded whereas for the Brownian like motion of $p(n)$, the $M(n)$ linearly grows in time. From this the asymptotic growth rate $K$ is defined by
 
\begin{equation}
K=\lim \limits_{N \to \infty}\frac{log~M(n)}{log~n}.
\end{equation}

Now the $K=0$ means the regular and $K=1$ means chaotic dynamics of the analyzed data. 

It is proven that this method is suitable for the time series data with low noise \cite{ref31}. Since the experimental time series are always has small amount of noise which is due to the inherent noise fluctuation present in the circuit itself. We perform the ``0-1 test" for the experimental data collected during the periodic and chaotic dynamics of the proposed circuit (Fig. 7). The set of `n'data points of the time series `$x_4$' after the initial transients vanished, is collected to the computer from the circuit using ``Agilent-U2531A" data acquisition module with the sampling rate of 500 kSa/s. The value of the constant `c' is randomly varied between ($0-\pi$). The calculated translation components ($p,s$), the mean square displacement $M(n)$ and the asymptotic growth rate $K$ for regular (period-1 limit cycle at $R15$ = 59.3~k${\mathrm{\Omega}}$) and chaotic time series (double band chaotic attractor at $R15$ = 54.6~k${\mathrm{\Omega}}$) are plotted in Fig. 15. These plots clearly differentiates the regular and chaotic motion of the circuit.

\section{Conclusions}
In this paper, the Chua's circuit with smooth, cubic functional memristor is experimentally designed and implemented. The difficulties on physical realization of memristor sub-circuit are bypassed using a circuit design based in electronic analogy. To avoid the inherent noise influence on variables oscillating on low amplitudes, the amplitude scaling is applied to original circuit equations such that the maximum amplitude of the oscillations is $\pm 5$V. This method of implementation is straight forward and it allows to reproduce all possible dynamics captured in computational simulations. The designed circuit is flexible and robust, presenting well comparable size in relation to other designs proposed in the literature. 
The experimental implementation provides time waveforms, phase portraits, and power spectra that present a good agreement with numerical simulation. Further, the numerical simulation also confirms the period doubling bifurcation scenario identified in the experiment. The regular and chaotic motion of the experimental data are clearly distinguished using ``0-1 test". In this context, the proposed analogue circuit can be used as a prototype model for memristor based circuit, consisting in a good option for the investigation of strange non-chaos, chaotic beats, nonlinear resonances and other associated behaviors.
   
\vskip 5pt
\section*{Acknowledgments}
 \vskip 5pt
The authors acknowledges the technical support of K. Suresh, S. Sabarathinam and P. Megavarna Ezhilarasu during various stages of this work. R.J. is supported by the University Grants Commission, India in the form of Research Fellowship in Science for Meritorious Students. The work of K.T. forms a part of a Department of Science and Technology, Government of India sponsored project grant no. SB/EMEQ-077/2013.

\begin{figure}[ht]
\begin{center}
\includegraphics[width=0.8\linewidth]{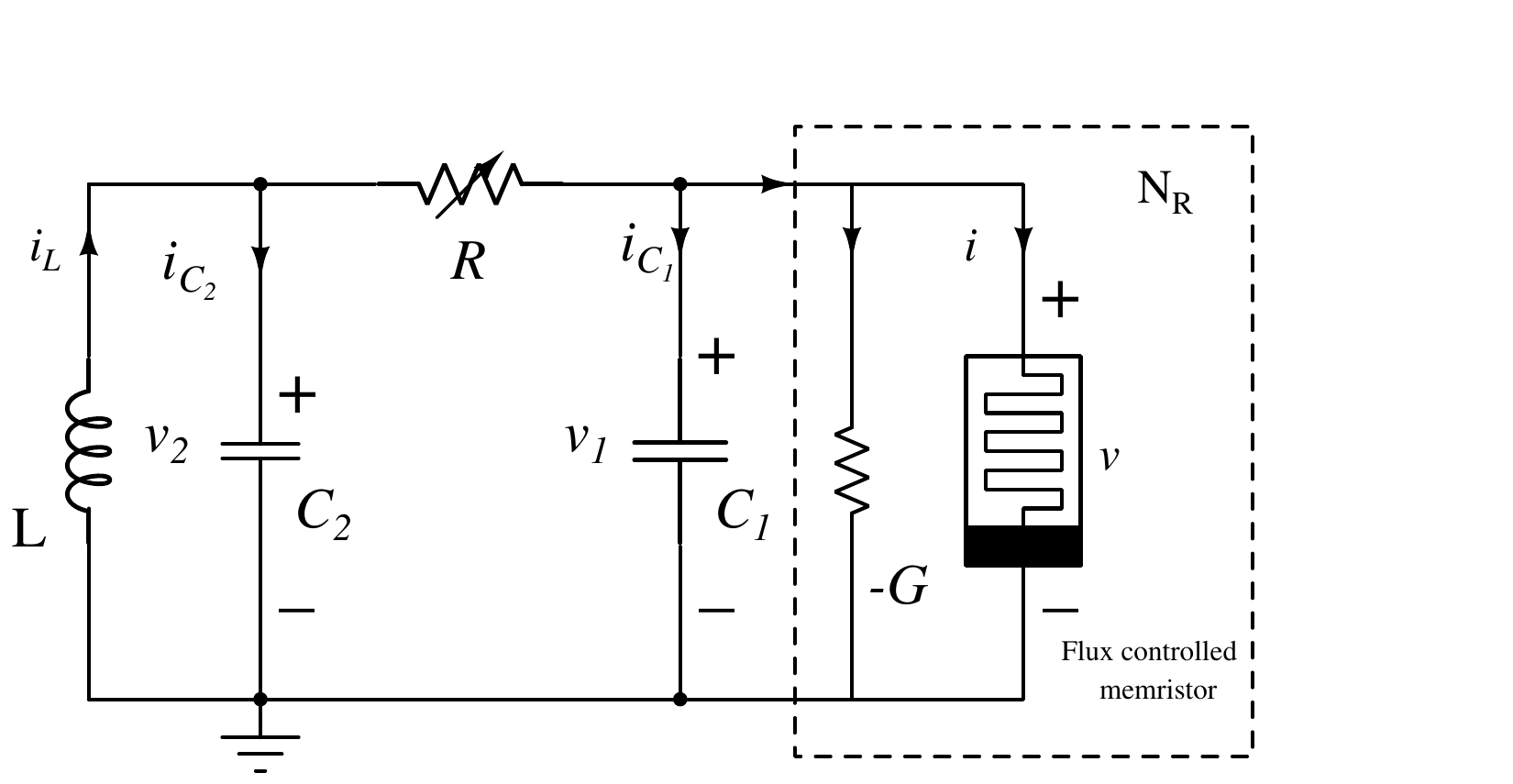}
\end{center}
\caption{The Chua's circuit with memristor ($N_R$).}
\end{figure}

\begin{figure}[ht]
\begin{center}
\includegraphics[width=1.0\linewidth]{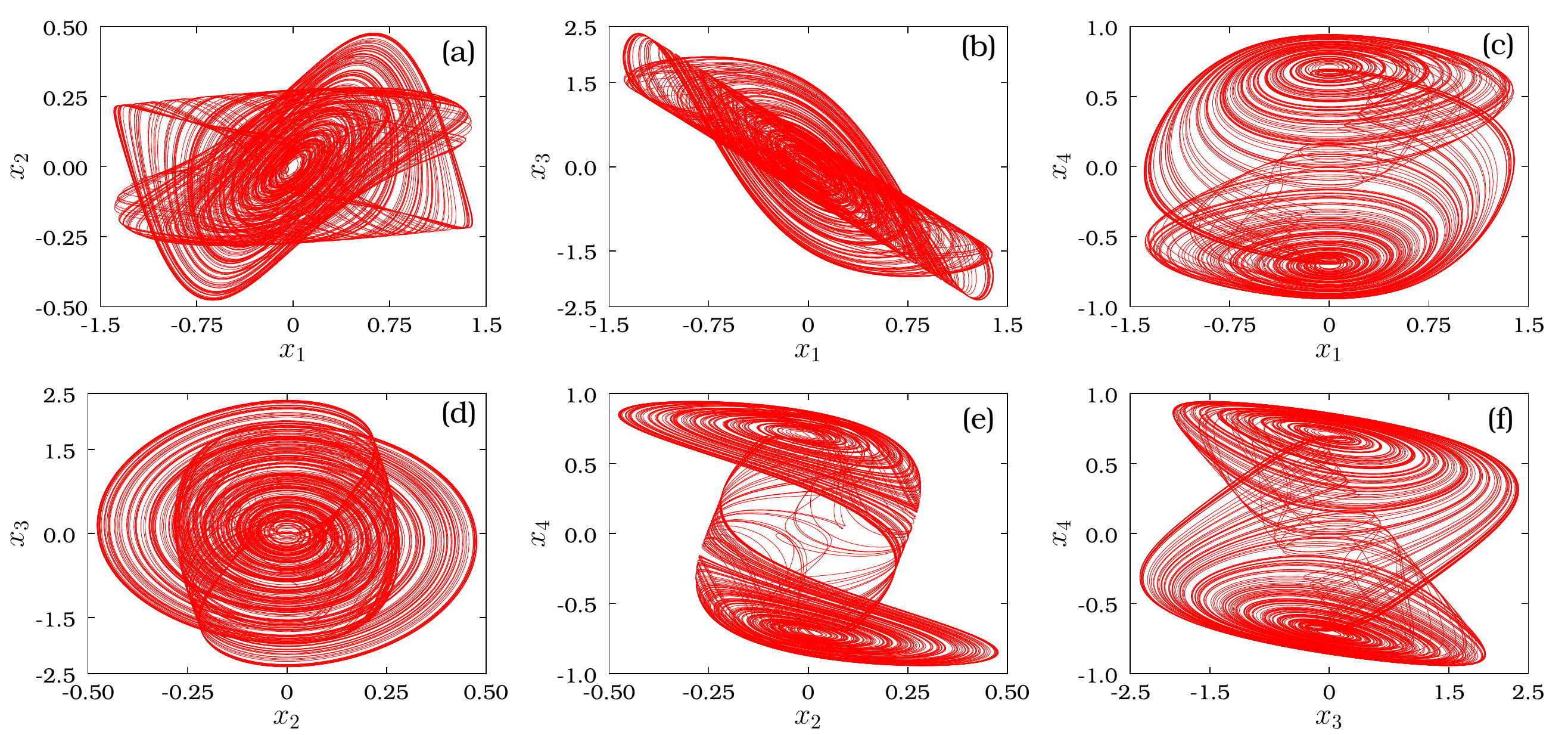}
\end{center}
\caption{(Color online) Different projections of phase portraits in the (a) ($x_1-x_2$), (b) ($x_1-x_3$), (c) ($x_1-x_4$), (d) ($x_2-x_3$), (e) ($x_2-x_4$) and (f) ($x_3-x_4$) planes obtained numerically from Eqns. 4. The parameters of the system are fixed as $\alpha$ = 9.8, $\beta$ = 100/7, $\gamma$ = 9/7, $a = $1/7 and $b = $2/7. The initial conditions are \{$x_1, x_2, x_3, x_4$\}=\{0, 0, 0.1, 0\}. }
\end{figure}

\begin{figure}[ht]
\begin{center}
\includegraphics[width=0.8\linewidth]{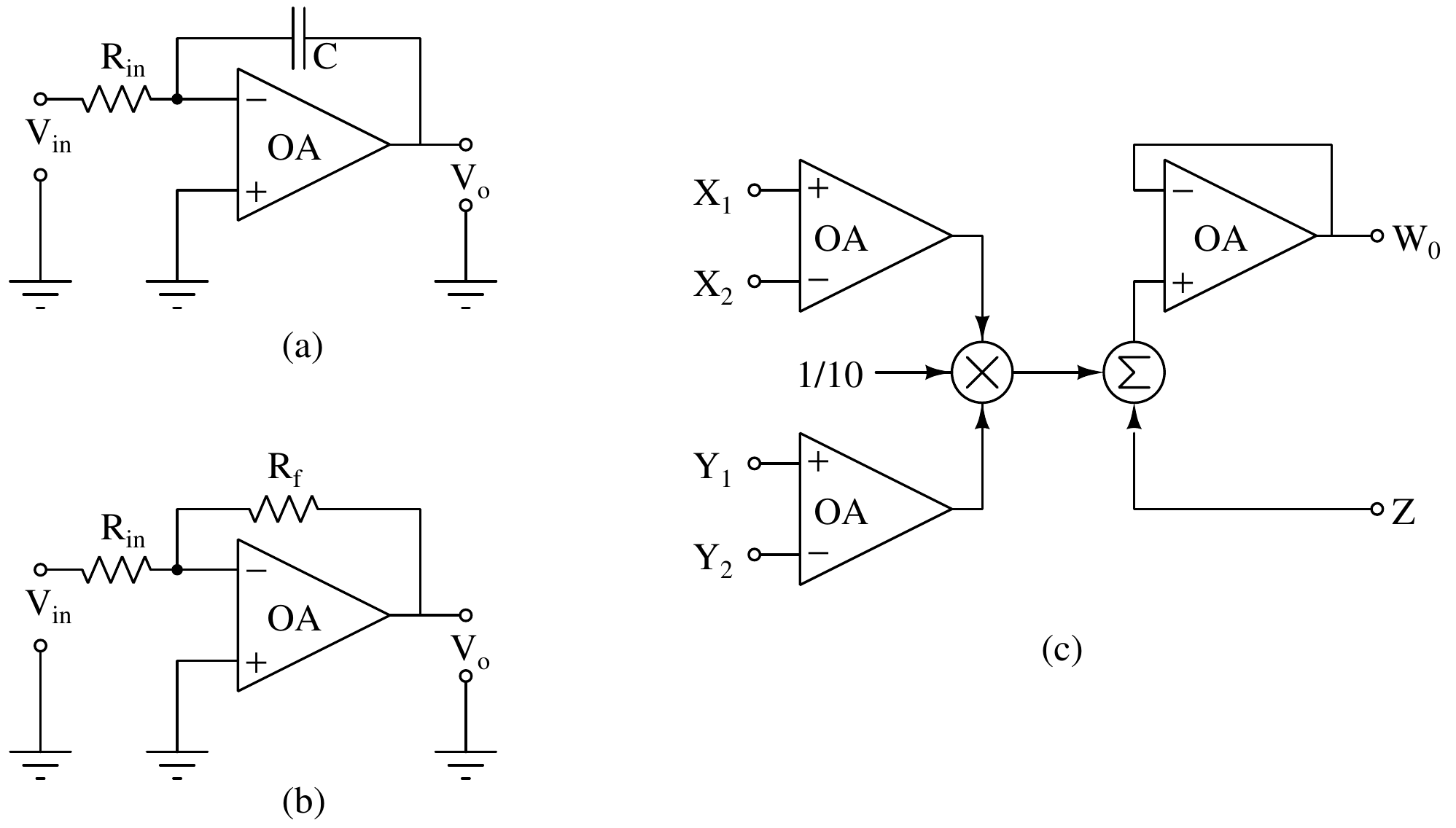}
\end{center}
\caption{The fundamental units of the analogue computation. Namely, (a) the integrator, (b) the inverting amplifier and (c) the schematic of an analog multiplier.}
\end{figure}

\begin{figure}[ht]
\begin{center}
\includegraphics[width=0.9\linewidth]{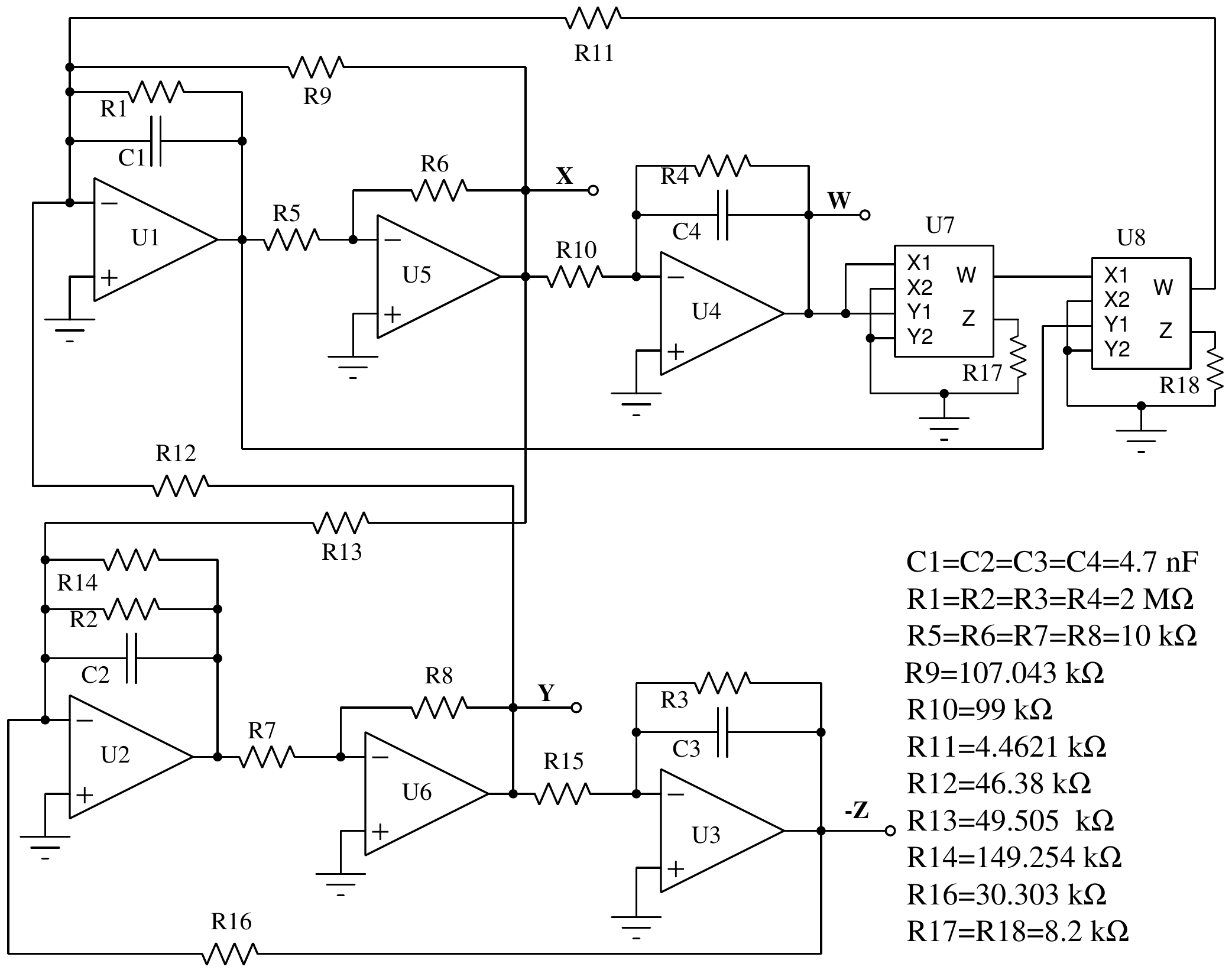}
\end{center}
\caption{The analogue circuit diagram of the memristor based Chua's circuit. The component values given in the diagram are used for PSpice simulation. For U1-U6, TL081 op-amps and for U7 and U8, AD633JN multiplier chips are used.}
\end{figure}

\begin{figure}[ht]
\begin{center}
\includegraphics[width=1.0\linewidth]{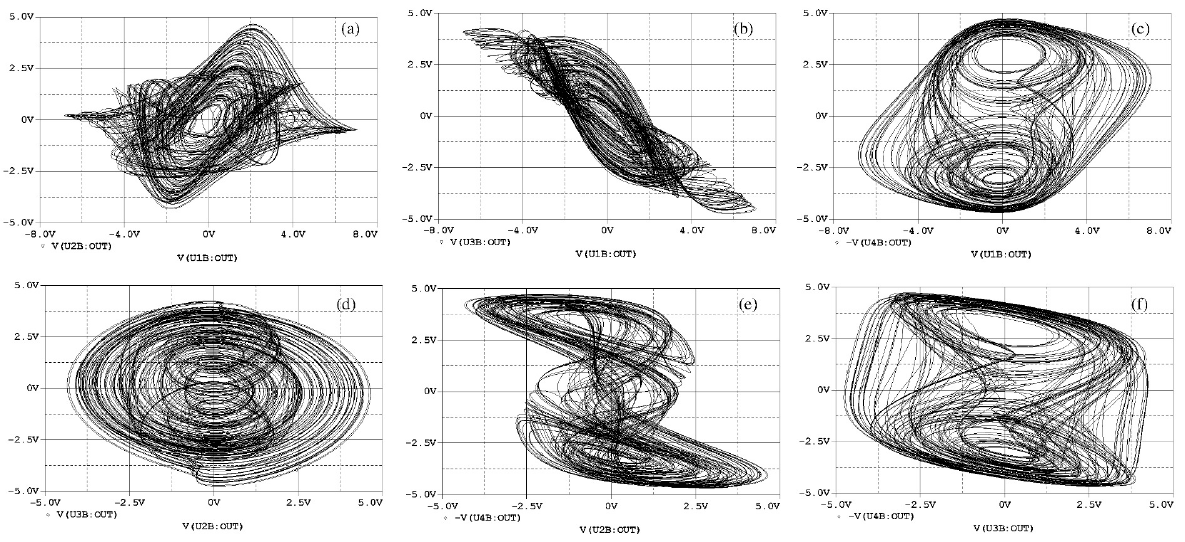}
\end{center}
\caption{Different projections of PSpice simulated phase portraits in the (a) ($v_1-v_2$), (b) ($v_1-v_3$), (c) ($v_1-v_4$), (d) ($v_2-v_3$), (e) ($v_2-v_4$) and (f) ($v_3-v_4$) planes.}
\end{figure}

\begin{figure}[ht]
\begin{center}
\includegraphics[width=0.8\linewidth]{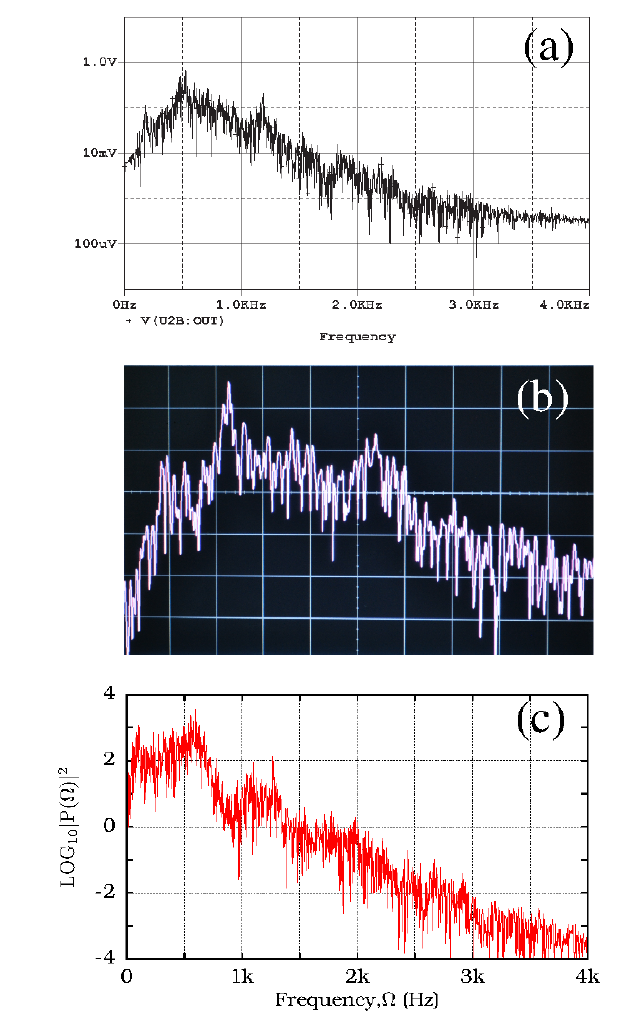}
\end{center}
\caption{(Color online) Fast Fourier transform spectrum of the variable $v_2$ for different simulation platforms with central operating frequency. (a) PSpice circuit simulation : 523 Hz, (b) Hardware circuit : 454 Hz, Scale: Horizontal axis = 200 Hz/div, Vertical axis = 10dB/div and (c) Numerical simulation : 532 Hz. }
\end{figure}

\begin{figure}[ht]
\begin{center}
\includegraphics[width=0.85\linewidth]{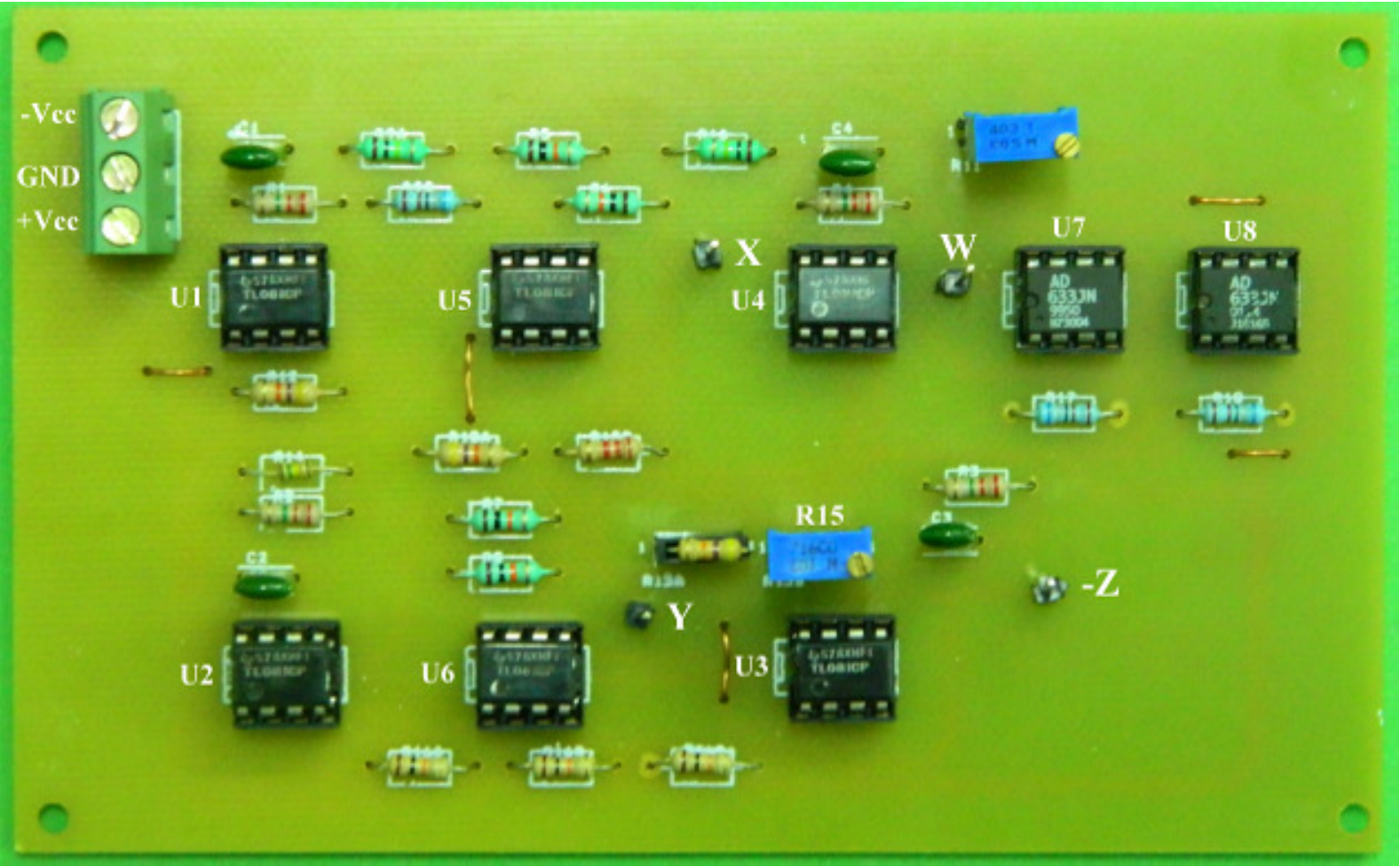}
\end{center}
\caption{(Color online) Experimental setup of the memristor based Chua's circuit. U1-U6 are TL081 op-amps, U7,U8 are AD633 mulitipliers. The nodes `X', `Y', `Z' and `W' are outputs of the circuit. The variable resistor ``R15" is used as a control parameter.}
\end{figure}

\begin{figure}[ht]
\begin{center}
\includegraphics[width=1.0\linewidth]{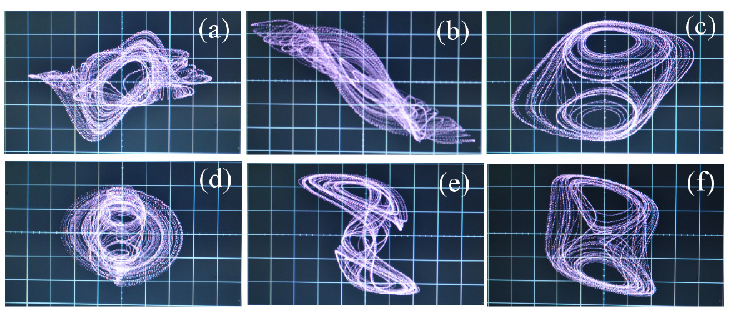}
\end{center}
\caption{(Color online) Different projections of phase portraits obtained from experiment in the (a) ($v_1-v_2$), (b) ($v_1-v_3$), (c) ($v_1-v_4$), (d) ($v_2-v_3$), (e) ($v_2-v_4$) and (f) ($v_3-v_4$) planes. Scale: Horizontal axis = 1.36 V/div, Vertical axis = 1.52 V/div. }
\end{figure}

\begin{figure}[ht]
\begin{center}
\includegraphics[width=0.8\linewidth]{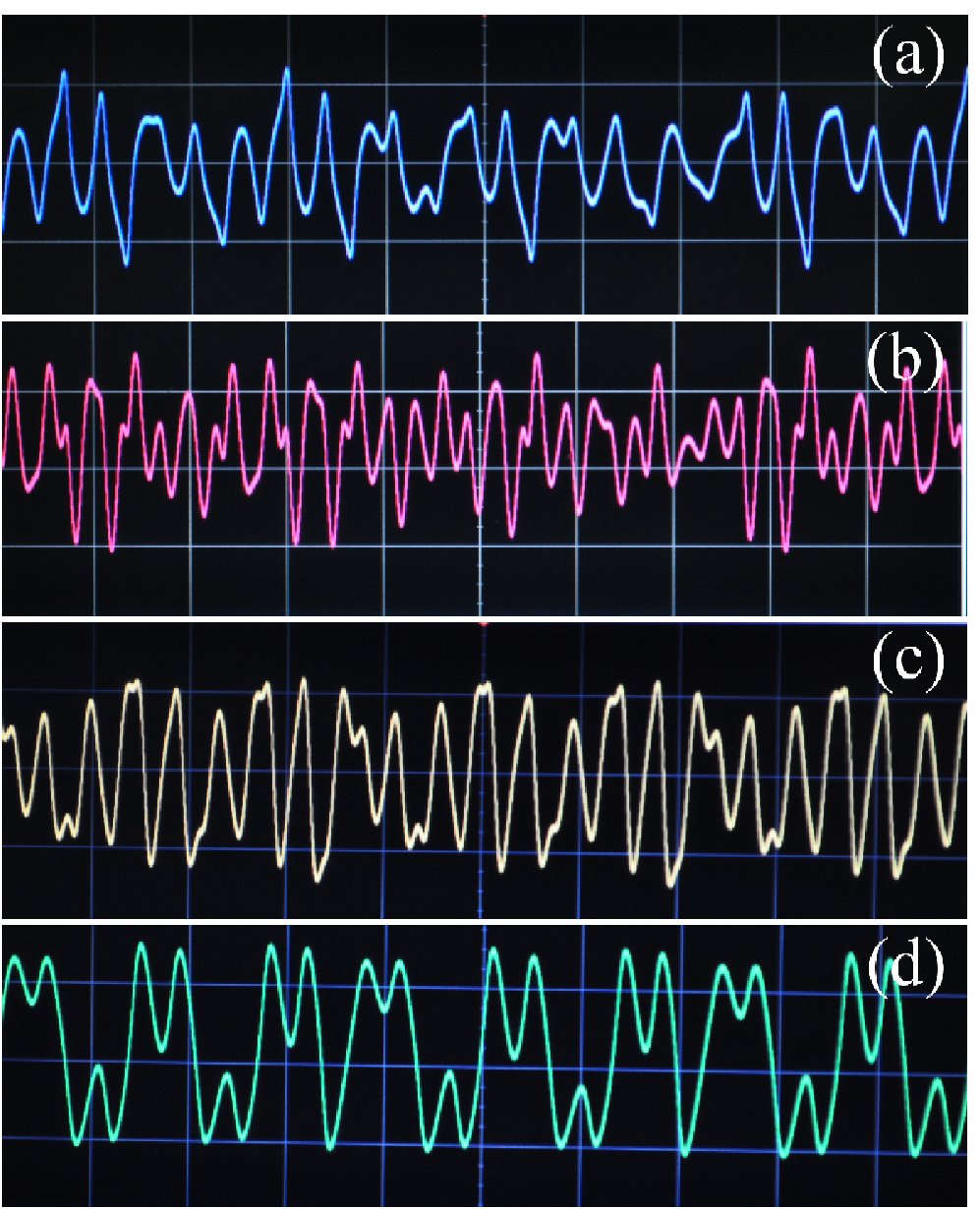}
\end{center}
\caption{(Color online) Experimental time waveform of the variables $v_1(t), v_2(t), v_3(t)$ and $v_4(t)$. Scale: Horizontal axis = 5 mS/div, Vertical axis = (a) 2.5 V/div, (b) 3.0 V/div, (c) 4.0 V/div and (d) 2.5 V/div.}
\end{figure}

\begin{figure}[ht]
\begin{center}
\includegraphics[width=0.8\linewidth]{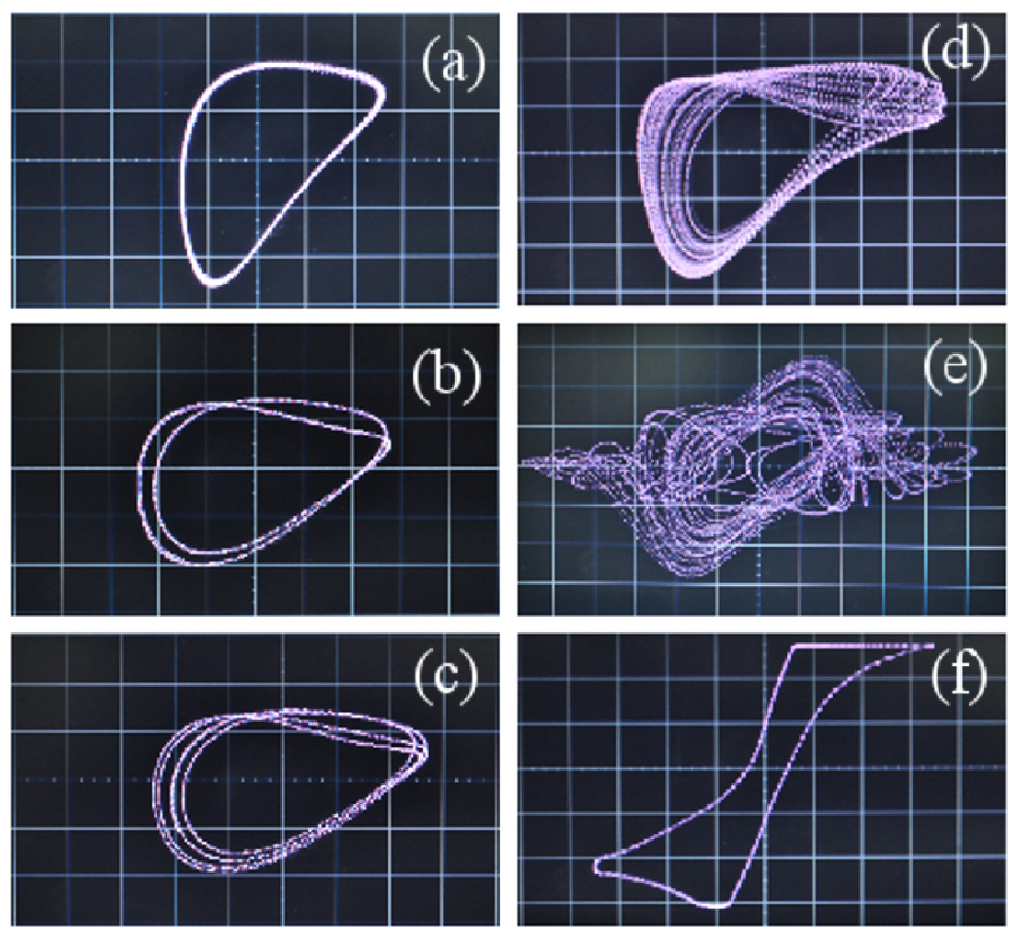}
\end{center}
\caption{(Color online) Period doubling scenario of experimental phase portraits in the ($v_1-v_2$) plane. (a) period-1 limit cycle ($R15 = $ 59.3 k${\mathrm{\Omega}}$), (b) period-2 limit cycle ($R15$ = 57.2 k${\mathrm{\Omega}}$), (c ) period-4 limit cycle ($R15$ = 55.9 k${\mathrm{\Omega}}$), (d) single band chaos ($R15$ = 55.1 k${\mathrm{\Omega}}$), (e) double band chaos ($R15$ = 54.6 k${\mathrm{\Omega}}$) and (f) saturated limit cycle ($R15$ = 53.1 k${\mathrm{\Omega}}$).}
\end{figure}

\begin{figure}[ht]
\begin{center}
\includegraphics[width=0.8\linewidth]{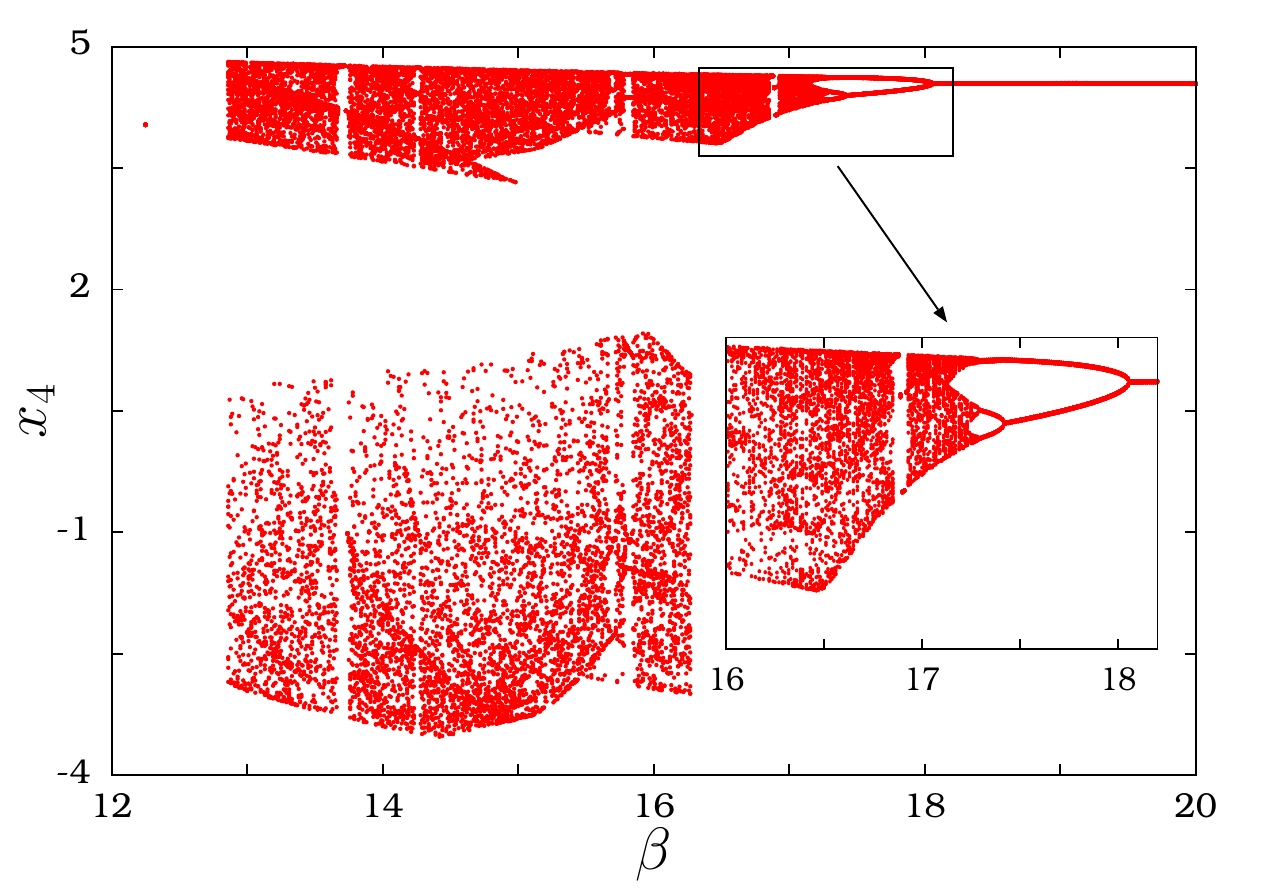}
\end{center}
\caption{(Color online) The one parameter diagram in the ($\beta-x_4$) plane obtained by numerical simulation. The period doubling sequence is clearly captured in the inset.}
\end{figure}

\begin{figure}[ht]
\begin{center}
\includegraphics[width=0.9\linewidth]{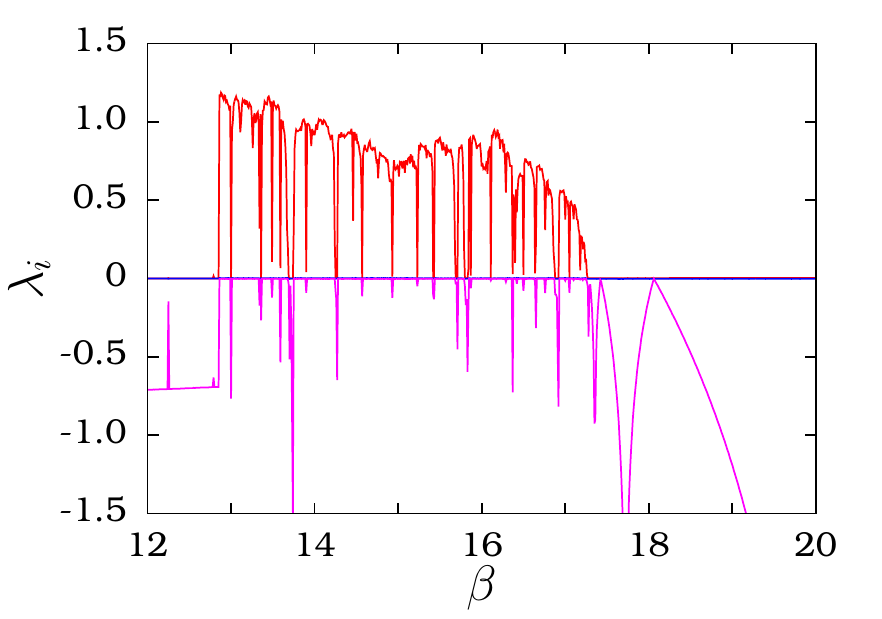}
\end{center}
\caption{(Color online) The Lyapunov exponents spectrum for first three exponents ($\lambda_i$ for $i = 1,2$ and $3$) are plotted in the ($\beta-\lambda_i$) plane. The fourth exponent is skipped to view the rest conveniently. The positive values in the $\lambda_1$ indicates the chaotic nature of the system. }
\end{figure}

\begin{figure}[ht]
\begin{center}
\includegraphics[width=1.0\linewidth]{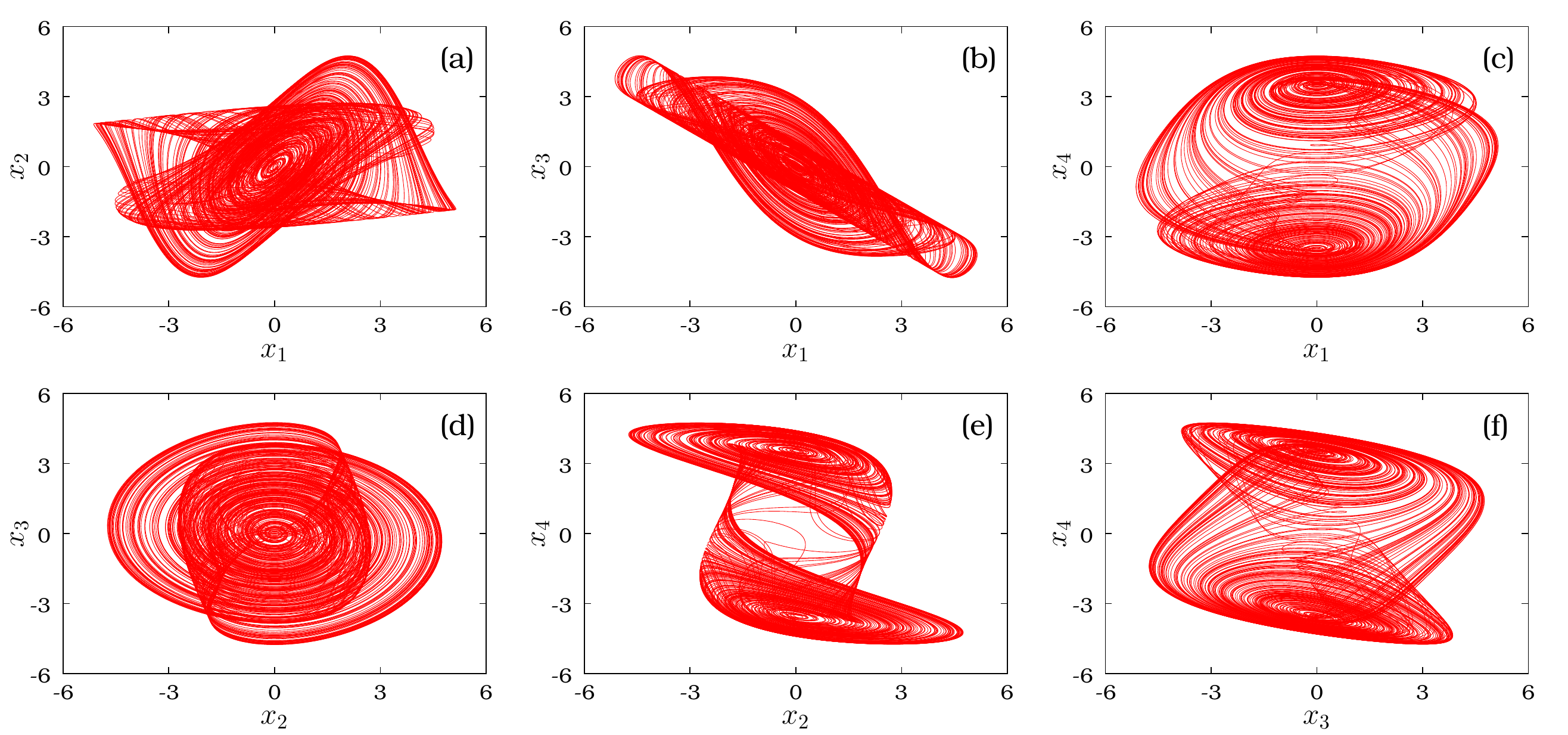}
\end{center}
\caption{(Color online) The numerical phase portraits of Eqns. (6) in the (a) ($x_1-x_2$), (b) ($x_1-x_3$), (c) ($x_1-x_4$), (d) ($x_2-x_3$), (e) ($x_2-x_4$) and (f) ($x_3-x_4$) planes. The parameters of the system are fixed as $\alpha$ = 9.8, $\beta$ = 100/7, $\gamma$ = 9/7, $a = $1/7 and $b = $2/7. The initial conditions are \{$x_1, x_2, x_3, x_4$\}=\{0, 0, 1.0, 0\}.}
\end{figure}

\begin{figure}[ht]
\begin{center}
\includegraphics[width=1.0\linewidth]{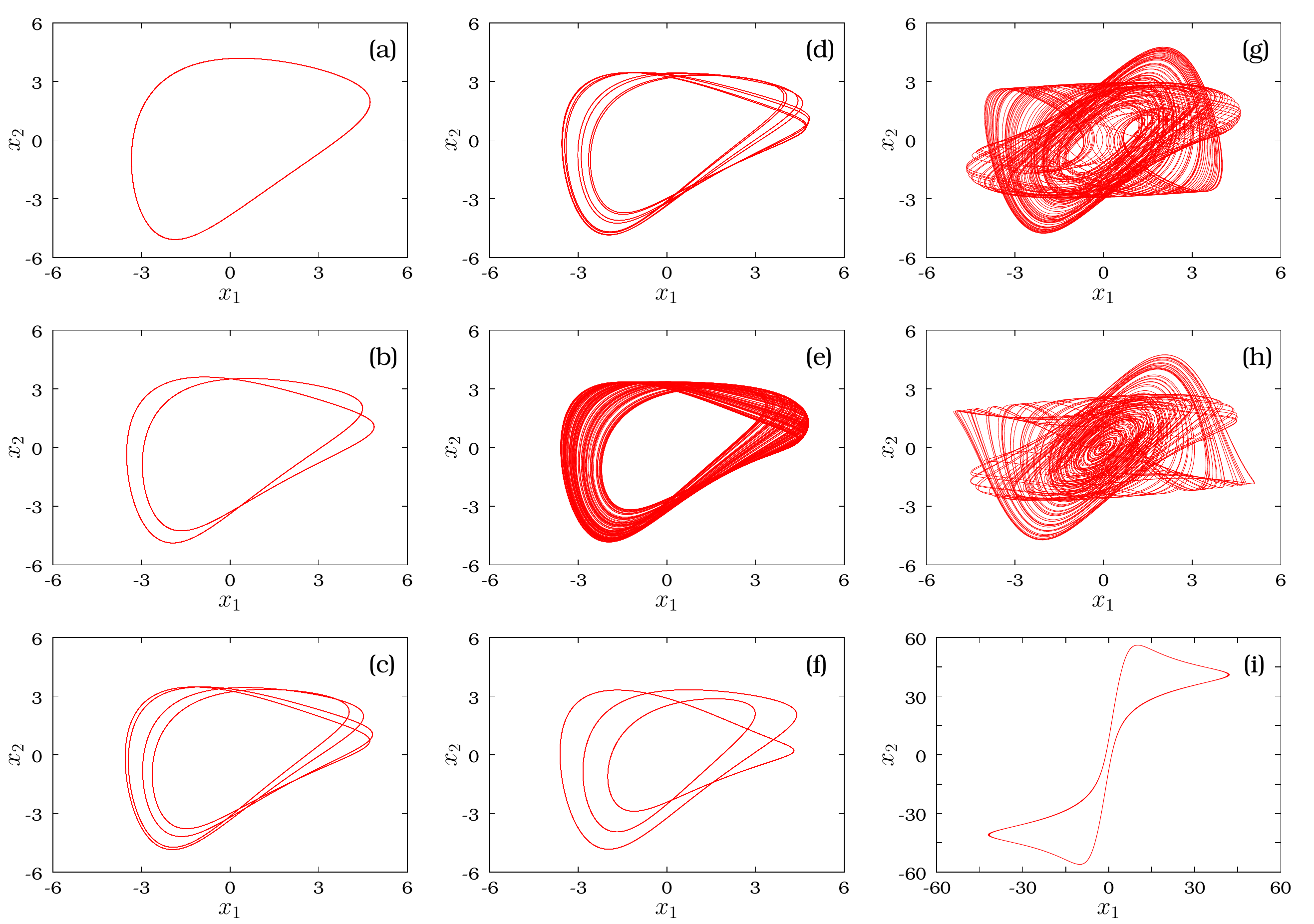}
\end{center}
\caption{(Color online) The numerical phase portraits of Eqns. (6) in the ($x_1-x_2$ for different values of $\beta$. (a) $\beta$ = 20, (b) $\beta$ = 17.7, (c) $\beta$ = 17.3, (d) $\beta$ = 17.268, (e) $\beta$ = 17 and (f) $\beta$ = 16.9. Rest of the parameters are same used for plotting Fig. 13.}
\end{figure}

\begin{figure}[ht]
\begin{center}
\includegraphics[width=1.0\linewidth]{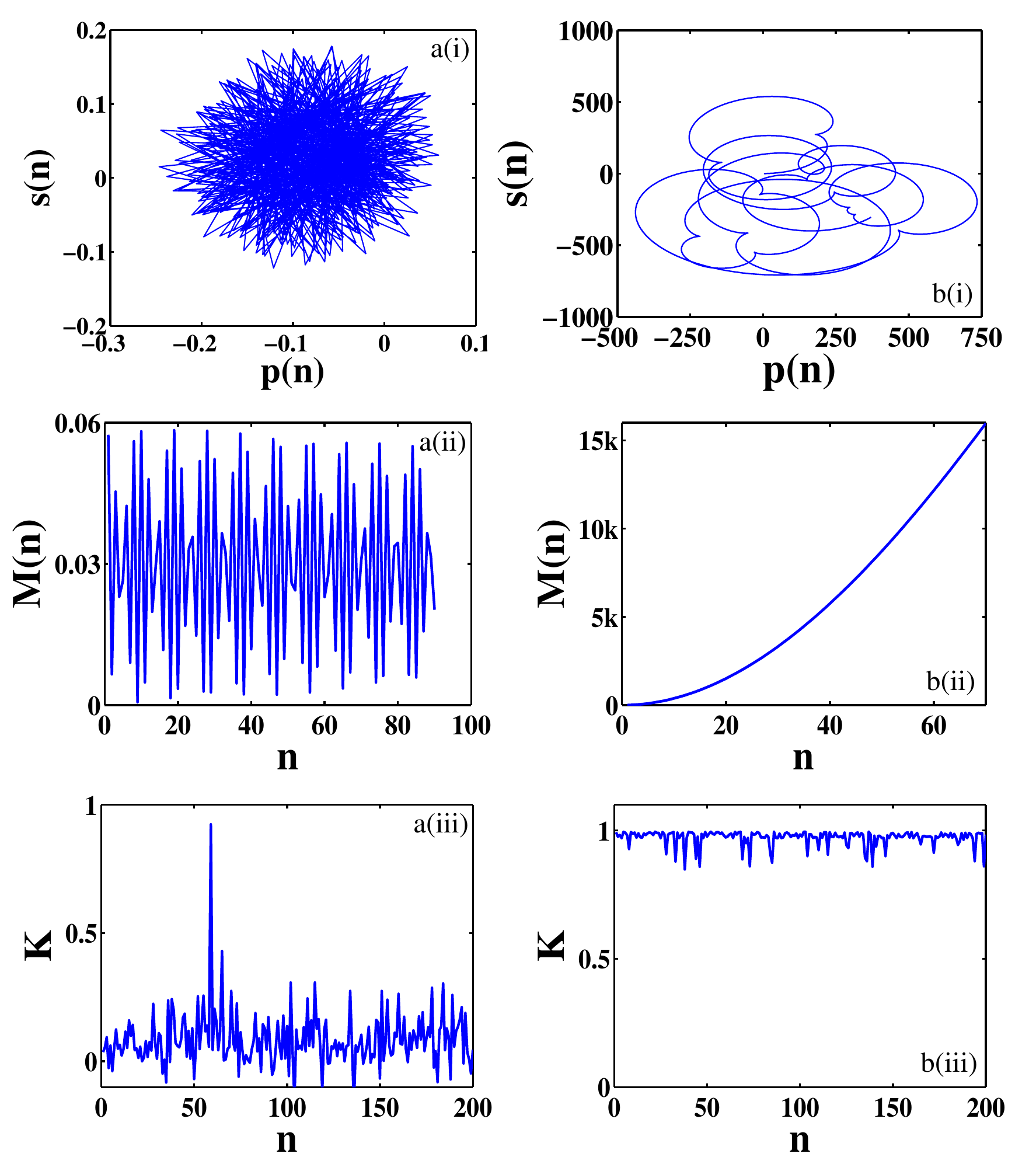}
\end{center}
\caption{(Color online) Application of the “0-1 test” to the experimental time series for (a) regular ($R15$ = 59.3~k${\mathrm{\Omega}}$) and (b) chaotic motion ($R15$ = 54.6~k${\mathrm{\Omega}}$): (i) Translation dynamics of ($p(n),s(n)$), (ii) mean square displacement $M(n)$ and (iii) asymptotic growth rate $K$.}
\end{figure}

\end{document}